\documentclass[final,5p,times,twocolumn]{elsarticle}
\biboptions{sort&compress}
\usepackage{isotope}
\usepackage[T1]{fontenc}
\usepackage{amsmath,bbold,amssymb,epsfig,feynmp,color,ifpdf}
\usepackage{nicefrac,exscale,multirow,times,txfonts}
\usepackage[colorlinks,linkcolor=blue,anchorcolor=purple,citecolor=blue]{hyperref}
\usepackage{mathrsfs}
\usepackage{graphicx}
\usepackage{dcolumn}
\usepackage{bm}
\usepackage{booktabs}
\usepackage[section]{placeins}

\makeatletter
\def\ps@pprintTitle{%
  \let\@oddhead\@empty
  \let\@evenhead\@empty
  \def\@oddfoot{} 
  \let\@evenfoot\@oddfoot
}
\makeatother

\usepackage{ulem}

\newcommand{\delete}{\bgroup\markoverwith{\textcolor{red}{\rule[0.5ex]{2pt}{1pt}}}\ULon}

\begin{document}
	\begin{frontmatter}
		
		\title{Impact of pion tensor force on alpha clustering in $^{20}$Ne}
		
		\author[first,second]{Zhao Jing Chen}
		\author[first,second]{Bao Yuan Sun*}
		\affiliation[first]{organization={MOE Frontiers Science Center for Rare Isotopes, Lanzhou University},
			city={Lanzhou},
			postcode={730000},
			country={China}}
		\affiliation[second]{organization={School of Nuclear Science and Technology, Lanzhou University},
			city={Lanzhou},
			postcode={730000},
			country={China}}

		\begin{abstract}
			The nuclear clustering, as a quantum phase transition phenomenon governed by strong interactions, exhibits characteristics that are highly sensitive to the specific features of nuclear forces. Here, we examine how nuclear deformation and tensor forces influence $\alpha$-cluster formation in light nuclei. The axially deformed relativistic Hartree-Fock-Bogoliubov model is utilized to investigate the clustering structure of the $^{20}$Ne nucleus, at both the ground state and the excited state with a superdeformed prolate. The nuclear binding energies and the canonical single particle levels are obtained at different quadruple deformation, and the role of tensor force embedded in the Fock diagram of $\pi$-pseudovector ($\pi$-PV) coupling is revealed. It is shown that the level branches from the degenerated spherical orbits at the deformed prolate case are enlarged due to the extra contribution from pion-exchanged tensor force. Correspondingly, the excitation energy in this superdeformed prolate state is reduced due to the noncentral tensor interaction, leading to a predicted value which is much closer to the referred threshold for the $2\alpha$ decay mode of $^{20}$Ne. Possible $\alpha$-clustering configurations in $^{20}$Ne are then characterized by examining the nucleonic localization function. Although the contribution to the ground state is relatively small, the density profile and nucleonic localization are significantly changed by the pion tensor force for the superdeformed prolate excited state, as further evidenced by characterising the level mixing in the spherical basis components. The results reveal the extra role of the tensor force, correlated to the evolved single-particle levels with nuclear deformation, in the formation and stability of nuclear clustering.
		\end{abstract}
		
		
	\end{frontmatter}
	
	The phenomenon of clustering occurs widely in the universe, from macroscopic galaxies to microscopic atomic nuclei. As early as the 1930s, Gamow and his colleagues suggested that $\alpha$-conjugated nuclei such as $^{12}$C, $^{16}$O, and $^{20}$Ne are composed of $\alpha$ particles~\cite{Gamow1930}. With the advancement of experimental techniques, the phenomenon of nuclear clustering has been observed in both light and heavy nuclei~\cite{BIJKER2020103735,RevModPhys.90.035004,RevModPhys.89.011002,FUNAKI201578,wei_clustering_2024,PhysRevLett.112.162501,PhysRevLett.131.212501}. Clustering phenomena typically emerge in deformed nuclear configurations~\cite{PhysRevLett.112.162501}, manifesting as excited states positioned near their corresponding decay thresholds~\cite{10.1143/PTPS.E68.464}. Although the mechanism of nuclear clustering is not yet fully understood, its origin is fundamentally tied to nuclear interactions. The emergence of clustering can be regarded as a quantum phase transition~\cite{natureebran_how_2012,PhysRevLett.117.132501}, which is potentially sensitive to the precise nucleon-nucleon interaction. Therefore, the study of clustering phenomena holds significance for the research of nuclear interactions.
	
	The $^{20}$Ne nucleus exhibits significant intrinsic quadrupole deformation and clear spatial localization in its density distributions.~\cite{PhysRevC.87.057309,ADACHI2021136411,PhysRevC.111.L021901,PhysRevLett.110.262501,PhysRevC.101.064308,naturezhou_5_2023}. These studies have provided valuable insights into the underlying mechanisms and phenomena associated with nuclear clustering. In previous work, various theoretical models have been developed to predict the $\alpha$-clustering phenomenon in $^{20}$Ne nuclei, including generator coordinate method~\cite{naturezhou_5_2023}, antisymmetrized molecular dynamics~\cite{PhysRevC.69.044319}, no-core shell model~\cite{PhysRevC.102.044608}, and covariant density functional theory~\cite{natureebran_how_2012,zhou_anatomy_2016}. As one of the typical representatives of covariant density functional theory, the relativistic mean field (RMF) theory with only Hartree approximation efficiently explores nuclear structure across the nucleon chart~\cite{P-GReinhard_1989,RING1996193,RevModPhys.75.121,VRETENAR2005101,MENG2006470}, successfully reproducing $\alpha$ clustering~\cite{natureebran_how_2012,PhysRevC.94.064323,PhysRevC.106.064330,PhysRevC.107.064605} and molecular states~\cite{PhysRevC.90.054329} in several nuclei.
	
	Compared to the RMF theory, the density dependent relativistic Hartree-Fock (RHF) theory~\cite{Long_ctp2022,LONG2006150} includes Fock terms that can explain important characteristics associated with $\pi$-pseudovector ($\pi$-PV) and $\rho$-tensor couplings. The RHF theory achieves a description of nuclear ground state properties at the same level as RMF theory. The inclusion of the Fock terms impacts significantly key nuclear properties, such as the equation of state of nuclear matter and neutron stars~\cite{PhysRevC.78.065805,JPGZhao_2015,PhysRevC.103.014304}, hypernuclear physics~\cite{PhysRevC.85.025806,PhysRevC.106.054311,Ding_2023}, halo phenomenon~\cite{PhysRevC.81.031302,PhysRevC.87.034311,GENG2024139036}, and new magicity~\cite{LI2014169,LI201697}. 
	
	In the context of the RHF approach, the tensor force is introduced naturally through the Fock diagram of meson exchange picture~\cite{PhysRevC.91.034326,CPCZong_2018,PhysRevC.98.034313}. As a crucial component of nuclear interactions, the tensor force has been revealed to play an important role in shell evolution~\cite{RevModPhys.92.015002,PhysRevLett.97.162501,PhysRevC.74.061303,COLO2007227,LONG2009428,steppenbeck_evidence_2013,wienholtz_masses_2013}, isospin excitations~\cite{PhysRevLett.105.072501,PhysRevC.85.034323}, $\beta$-decays~\cite{BAI200928,PhysRevC.83.064306,PhysRevLett.110.122501} and symmetry energy~\cite{PhysRevC.91.025802,JPGZhao_2015}. In the RHF framework, tensor force mediated by $\pi$-PV critically affects the properties of spin-unsaturated nuclei~\cite{ELLong_2008,PhysRevC.87.047301,LONG2009428,LI201697}. When considering the $\pi$-PV coupling, the RHF model can improve the description of nuclear binding energies and potentially has a significant impact on nuclear shape evolution and polarization~\cite{PhysRevC.101.064302,GENG2024139036}. These findings underscore the importance of further investigating $\pi$-PV coupling and its tensor force component in understanding the formation and evolution of clustered structures.
	
	In this work, we employ the axially deformed symmetric relativistic Hartree-Fock-Bogoliubov (D-RHFB) model~\cite{PhysRevC.105.034329} to investigate the role of the tensor force in the clustering physics of $^{20}$Ne, focusing on both its ground state and the highly prolate deformed excited state. By isolating the tensor force contributions from $\pi$-PV coupling, we aim to extract its impact on the excitation energy and localization of $\alpha$ clustering in the largely deformed state. Using nucleonic localized functions (NLF)~\cite{PhysRevC.83.034312}, we extract clustering information to explore how the tensor force influences nuclear structure evolution and clustering phenomena in $^{20}$Ne, providing deeper insights into the role of tensor force in shaping nuclear dynamics.
	
	We choose the RHF Lagrangians PKO1 and PKO3 to address the effects of $\pi$-PV coupling on nuclear clustering~\cite{ELLong_2008,LONG2006150}. For comparison, the RHF Lagrangian PKO2 without $\pi$-PV and the RMF Lagrangian PKDD without Fock terms~\cite{PhysRevC.69.034319} are selected as well. Calculations are performed using the D-RHFB program based on the Dirac-Woods-Saxon basis~\cite{PhysRevC.105.034329}. The maximum value of magnetic quantum numbers $m$ is taken as $m_{\max}^+= 13/2$ for positive parity, and $m_{\max}^-= 15/2$ for negative parity. The number of $\kappa$ blocks considered in expanding the Dirac spinor set as $K_{m_{\max}}= 4$. The quadrupole deformation $\beta$ is defined as follows
	\begin{equation}
		\beta=\frac{\sqrt{5\pi}Q_2}{3R_0^2A},\quad Q_{2}={\frac{4\sqrt2\pi}{3}}\int\rho_{b}^{2}(r)r^{4} dr,
	\end{equation}
	where $\rho_b$ denotes the baryon density and $R_0=1.2A^{1/3}$ with $A$ the mass number. More details can be found in Ref.~\cite{PhysRevC.101.064302}. In the pairing channel, we utilize the finite-range Gogny force D1S~\cite{BERGER198423} as the pairing force, which naturally converges with the configuration space. We perform a quantitative analysis of the tensor force component in the $\pi$-PV coupling by subtracting the corresponding contributions in the mean field following the formula below
	\begin{equation}
		\mathscr{H}_{\pi-PV}^{T}=-\frac{1}{2}\left[\frac{f_{\pi}}{m_{\pi}}\bar{\psi}\gamma_{0}\Sigma_{\mu}\vec{\tau}\psi\right]_{1}\cdot\left[\frac{f_{\pi}}{m_{\pi}}\bar{\psi}\gamma_{0}\Sigma_{\nu}\vec{\tau}\psi\right]_{2}D_{\pi-PV}^{T,~\mu\nu}(1,2),\label{eq:4}
	\end{equation}
	where $D_{\pi-PV}^{T,~\mu\nu}(1,2)$ represents the propagator, $\Sigma^{\mu}=(\gamma^{5},\bm{\Sigma})$ is the relativistic spin operator, and $\vec{\tau}$ denotes the isospin operator of the nucleon field $\psi$. Eq.~(\ref{eq:4}) can be reduced to the nonrelativistic form of pion tensor force equivalently, see Ref.~\cite{PhysRevC.91.034326,PhysRevC.98.034313} for more details.
	
	To probe possible clustering information within atomic nuclei, it is necessary to employ tools such as the nucleon localization function. The key ingredient of the NLF is the leading term $Z_{q\sigma}(\vec r)$ of the Taylor expansion of the same-spin $(\sigma)$ and same-isospin $(q)$ conditional pair probability
	\begin{equation}
		Z_{q\sigma}(\vec{r})\equiv\tau_{q\sigma}(\vec{r})\rho_{q\sigma}(\vec{r})-\frac14[\vec{\nabla}\rho_{q\sigma}(\vec{r})]^2-\vec{j}_{q\sigma}^{~2}(\vec{r}),
	\end{equation}
	where $\rho_{q\sigma}$, $\tau_{q\sigma}$, $\vec{\nabla}\rho_{q\sigma}$ $\textrm{~and~}\vec{j}_{q\sigma} $ are the nucleon density, kinetic energy density, density gradient and current density. In the present case with static and time-reversal symmetry, the current density $\vec{j}_{q\sigma}$ vanishes. In D-RHFB model, then the rest of these quantities can be calculated by the canonical single-particle wave functions $\psi_{\nu\pi m,\sigma}$~\cite{PhysRevC.105.034329}. After normalized by $\tau_{q\sigma}^{\mathrm{TF}}\rho_{q\sigma}$, with $\tau_{q\sigma}^{\mathrm{TF}}$ denoting the Thomas-Fermi kinetic energy density
	\begin{equation}
		\tau_{q\sigma}^{\mathrm{TF}} = \frac{3}{5}(6\pi^2)^{2/3}\rho_{q\sigma}^{5/3} \equiv \rho_{q\sigma}^{5/3}/a,
	\end{equation}
	the NLF is then given by
	\begin{equation}
		C_{q\sigma}(\vec{r})=\left[1+\left(\frac{aZ_{q\sigma}(\vec{r})}{\rho_{q\sigma}^{8/3}(\vec{r})}\right)^2\right]^{-1}.
	\end{equation}
	When the conditional pair probability is low, indicating a high degree of nucleon localization at a specific position, the values $Z_{q\sigma}(r) = 0$ and $C_{q\sigma}(r) = 1$ are observed. In particular, in a nucleus where the neutron number equals to the proton number ($N = Z$), $C_{q\sigma}(r) = 1$ signifies a complete spatial overlap among the four possible nucleonic states defined by their isospin and spin configurations: $(q\sigma) = (n\uparrow, n\downarrow, p\uparrow, p\downarrow)$. This scenario is termed as a pure four-nucleon spatial overlap, taken as a necessary condition of $\alpha$ clustering inside a nucleus.
	
	In order to predict the existence of $\alpha$ clusters in atomic nuclei, a geometric condition beyond the NLF should also be introduced. The location where $\alpha$ clusters appear should exhibit a rapid decrease in density on the typical scale of $\alpha$ particles. Therefore, one could introduce the featured $F$ factor~\cite{PhysRevC.106.064330},
	\begin{equation}
		F(\vec r) = \frac{{{\rho _{\max }} - \rho (\vec r)}}{{{\rho _{\max }}}},
	\end{equation}
	where $\rho_{\max}$ denotes the peak density within the nucleus. A value of $F(\vec r)$ near one signifies a significant reduction in density, pointing to a depletion zone. Thus, it is anticipated that $F$ factor will be near one at the boundaries of the $\alpha$ cluster, marking the areas where the density experiences a sharp decline.
	
\begin{figure}[t]
		\centering
		\includegraphics[width=\columnwidth]{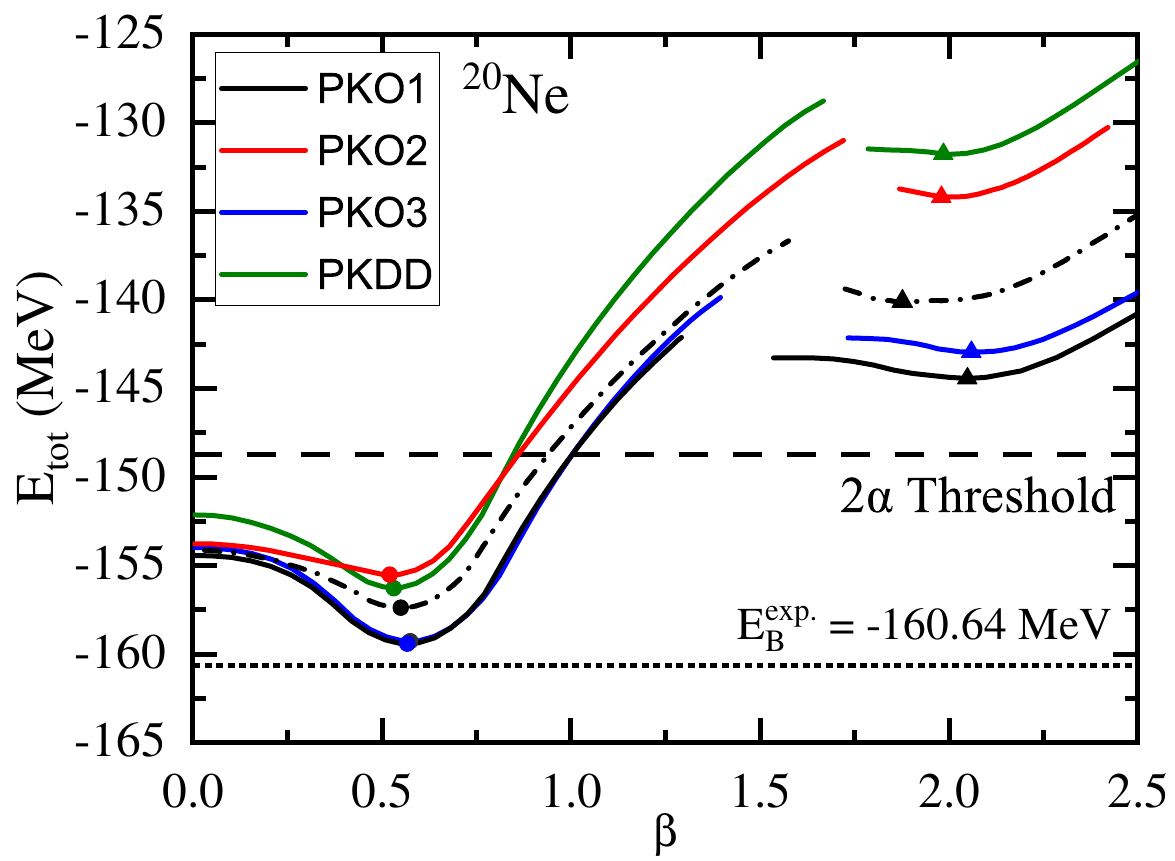}
		\caption{Binding energies as a function of the quadruple deformation $\beta$ for $^{20}$Ne. The results are calculated by the D-RHFB model with Lagrangians PKO1, PKO2, PKO3 and PKDD. The circles represent the ground state (G.S.) of $^{20}$Ne, while the triangles denote the excited state (E.S.). The experimental binding energy $E_B^{\textrm{exp.}}$ for the ground state of $^{20}$Ne is taken from Ref. \cite{CPCKondev_2021}, and the referred threshold for the two-alpha decay mode of $^{20}$Ne is shown by the dashed line \cite{10.1143/PTPS.E68.464}. For comparison, the result with PKO1 but excluding tensor force component of $\pi$-PV contributions from the nucleon self-energy (according to Eq.\ref{eq:4}) is given by the dash-dotted line as well.}
		\label{fig:1}
	\end{figure}

	We first examine the ground state (G.S.) properties of $^{20}$Ne. Fig.\ref{fig:1} displays the binding energies ($E_B=-E_{tot}$, in MeV) as a function of quadrupole deformation parameter $\beta$ calculated with the Lagrangians PKO$i$ and PKDD. The ground state of $^{20}$Ne exhibits a quadrupole deformation of $\beta \approx 0.55$, where the PKO1 and PKO3 functionals predict binding energies consistent with experimental data~\cite{CPCKondev_2021}. Notably, PKO2 and PKDD show an $E_B$ deficit of approximately 4 MeV compared to PKO1. A secondary minimum emerges at $\beta \approx 2.05$, corresponding to a largely deformed prolate configuration identified as an excited state (E.S.). At this state, significant discrepancies in $E_B$ are observed among the functionals. PKO1 gives the highest $E_B$ (144.4 MeV) while PKDD yields the lowest (131.7 MeV), indicating a 12.7 MeV energy difference. As PKO1 include extra pion-exchanged contribution from the Fock diagram, the disparity then suggests an essential role of $\pi$-PV coupling in the nuclear stability at large deformation. Furthermore, the PKO1 result approaches the theoretically predicted 2$\alpha$-cluster decay threshold of 11.89 MeV~\cite{10.1143/PTPS.E68.464}, implying the appearance of two distinct $\alpha$ clustering structures.
	
	\begin{table*}
		\caption{Contributions to the total energy $E_{tot}$ from various channels for both ground state (G.S.) and excited state (E.S.) of $^{20}$Ne, calculated by the D-RHFB model with Lagrangians PKO1, PKO2, PKO3 and PKDD. In detail, $E_k$ denotes the kinetic energy of nucleons, $E^D$ and $E^E$ are Hartree and Fock terms of the meson-exchange potential, and $E_\mathrm{other}$ includes those from the photon field, the pairing energy and the center-of-mass correction. All units are in MeV.}
		\label{table:1}
		\centering
		\begin{tabular}{lcrrrrrrrr}
			\toprule
			\multicolumn{2}{c}{} & \multicolumn{2}{c}{PKO1} & \multicolumn{2}{c}{PKO2} & \multicolumn{2}{c}{PKO3} & \multicolumn{2}{c}{PKDD} \\
			\hline
			&        & E.S. & G.S.    & E.S. & G.S. &  E.S. & G.S. &  E.S. & G.S. \\
			\hline
			$E_{tot}$ &&$-144.411$     &$-159.418$ & $-134.189$ &  $-155.527$&$-142.943$&	$-159.269$&$-131.745$&$-156.268$\\
			$E_k$ &&$318.274$ &$276.712$&$315.673$& $282.835$&$316.170$&	$276.289$&$297.052$&$264.013$\\
			
			$E^{D}$&${\sigma}$-S&$-2049.490$	&$-2015.911$	&$-1970.838$	&$-1998.279$	&$-2022.919$	&$-1986.927$&	$-2603.636$&	$-2687.157$\\
			
			&${\omega}$-V&	$1512.385$&	$1486.855$&$	1416.716$	&$1441.317$	&$1488.997$&	$1463.146$	&$2164.100$&	$2252.515$\\
			&${\rho}$-V&	$0.005$	&$0.005$	&$0.007$	&$0.005$	&$0.004$	&$0.005$	&$0.020$&$	0.022$\\
			$E^{E}$&${\sigma}$-S&$467.182$	&$459.647$	&$448.883$	&$454.843$	&$461.172$	&$453.013$&	$-$	&$-$\\
			
			&${\omega}$-V&	$-325.363$&	$-309.870$&	$-305.352$	&$-299.785$&$-321.302$	&$-305.673$&	$-$	&$-$\\
			&${\rho}$-V&	$-50.384$	&$-47.830$	&$-45.970$	&$-46.844$	&$-40.950$	&$-41.807$	&$-$	&$-$\\
			
			&${\pi}$-PV&	$-23.929$	&$-19.217$	&$-$	&$-$	&$-31.095$&$-27.482$	&$-$	&$-$\\
			$E_\mathrm{other}$&&	$6.910$	&$10.191$	&$6.696$	&$10.362$	&$6.982$&$10.167$	&$10.719$	&$14.342$\\
			\bottomrule
		\end{tabular}
	\end{table*}
	
	To elucidate the reasons for binding energy differences under various energy functionals, Table.\ref{table:1} summarizes the contributions of different meson-nucleon couplings to the $E_{tot}$ within the meson-exchange picture. Analysis of coupling channels reveals that the isoscalar channel modestly decreases $E_{tot}$, whereas the isovector channels, particularly via the $\pi$-PV coupling, significantly reduce it. This ultimately results in significantly lower total energies for PKO1 and PKO3 compared to other energy functionals, highlighting the role of $\pi$-PV coupling in affecting the stability of nuclei.
	
\begin{figure}[t]
		\centering
		\includegraphics[width=\columnwidth]{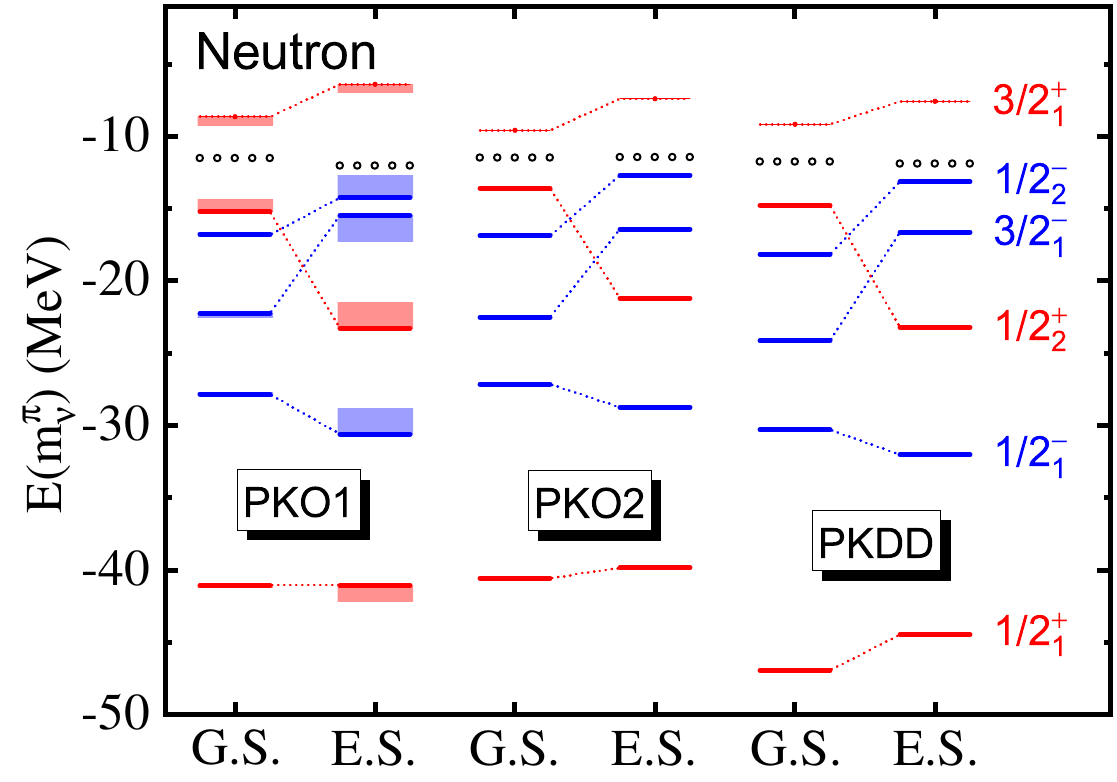}
		\caption{Neutron canonical single particle energies $E(m_{\nu}^\pi)$ for both G.S. and E.S. of $^{20}$Ne, calculated by the D-RHFB model with Lagrangians PKO1, PKO2 and PKDD. The positive(negative) parity levels are plotted by the red(blue) bars, respectively. For PKO1, the shadowed area illustrate the contribution due to extra tensor force of $\pi$-PV coupling in Fock terms.}
		\label{fig:2}
	\end{figure}

	The influence of $\pi$-PV coupling extends to single-particle spectra, as evidenced in Fig.\ref{fig:2} displaying neutron $E(m_{\nu}^\pi)$ levels for PKO1, PKO2, and PKDD at ground/excited states (with analogous proton spectra omitted due to similar trends), where PKO3 behavior aligns with PKO1 and is thus excluded. The notation $m_{\nu}^\pi$ is introduced to denote the single-particle orbits, where $\nu$ represents the principal quantum number, $m$ the magnetic quantum number, and $\pi$ the parity. Driven by quadrupole deformation evolution, single-particle levels undergo significant restructuring. At ground state, the single-particle level near the Fermi surface is ${1/2}_2^+$. With quadrupole deformation evolution in excitation, the ${1/2}_2^+$ becomes deeply bound, and the level closest to the Fermi surface becomes ${1/2}_2^-$. The similar phenomenon occurs in $^{11}$Be~\cite{GENG2024139036}, which is a characteristic feature of Nilsson energy levels under large quadrupole deformations. PKO1 exhibits a larger energy gap between the $3/2_1^-$ and $1/2_2^+$ orbitals than PKO2 and PKDD. This energy difference arises from enhanced splitting of degenerate spherical orbits under prolate deformation, significantly influenced by the $\pi$-PV coupling~\cite{PhysRevC.101.064302}. Ultimately, this facilitates $N=6$ shell closure formation. Shifts in single-particle levels and enhanced spin-orbit splitting demonstrate systematic modifications to single-particle dynamics by $\pi$-PV coupling in deformed nuclei.
	
	To isolate the role of the tensor force component, we performed constrained calculations in the D-RHFB program with Lagrangians PKO1 excluding tensor force component of $\pi$-PV contributions from the nucleon self-energy (PKO1 w.o. $V_{\pi-PV}^T$). The resulting single-particle energy shifts relative to full PKO1 are shown as shaded regions in Fig.\ref{fig:2}. According to the calculation results of PKO1 without $V_{\pi-PV}^T$, the shell closure structure of $N=6$ has weakened, while the value of each single-particle level returns to a value comparable to those obtained with the Lagrangians PKO2 and PKDD. This impact similarly manifests in binding energies, the binding energy as a function of $\beta$ using PKO1 without $V_{\pi-PV}^T$ is represented by the black dashed-dotted line in Fig.\ref{fig:1}. The energy minimum of E.S. appears near $\beta \approx 1.8$ for PKO1 without $V_{\pi-PV}^T$, compared to the minimum near $\beta \approx 2.0$ for Lagrangian PKO1. It can be found that the tensor force component of $\pi$-PV provides an additional attractive effect, which further reduces the energy of $^{20}$Ne and allows it to remain stable with larger prolate deformation for PKO1.
	
	\begin{figure*}[t]
		\centering
		\includegraphics[width=2\columnwidth]{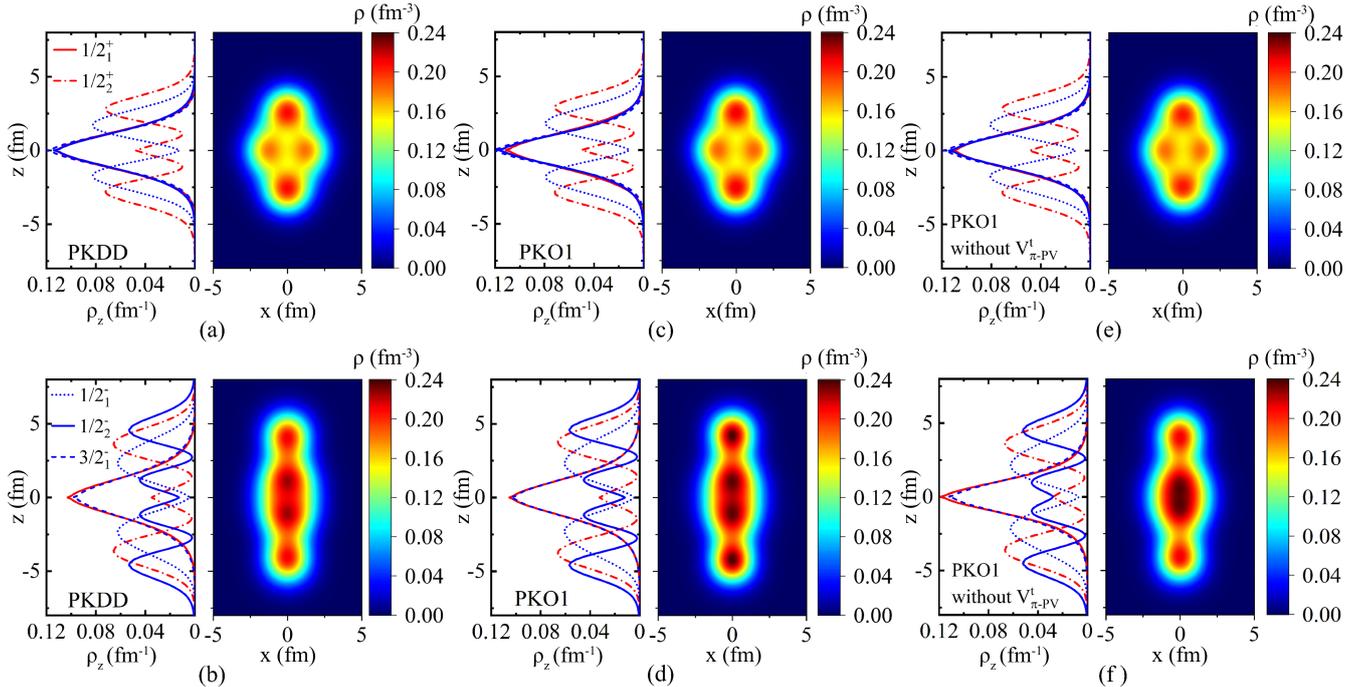}
		\caption{Figures (a) and (b) are calculated by the D-RHFB model with Lagrangians PKDD, figures (c) and (d) are calculated with the Lagrangians PKO1, while figures (e) and (f) show the results of the Lagrangians PKO1 with tensor force component of $\pi$-PV contribution removed from the nucleon self-energy (PKO1 w.o. $V_{\pi-PV}^T$). The top row displays the results of ground state, while the bottom row shows excited state. The left half of each sub-figure shows the distribution of the single-particle level density along the $z$-axis $(\rho_z)$ for the $^{20}$Ne, while the right half shows the contour plot of the density distribution $(\rho)$ of the $^{20}$Ne. The tensor force promotes cluster dissociation by adjusting the spatial distribution of single-particle orbitals (the $1/2^-_2$ expands outward).}
		
		\label{fig:3}
	\end{figure*}
	
	The differences in the Lagrangians of the models also impact predictions of the density distributions in $^{20}$Ne. Fig.\ref{fig:3} shows the density distribution along the $z$-axis $(\rho_z)$ for the single-particle level $m_{\nu}^\pi$ and the contour plots of the total density distributions for $^{20}$Ne, where $\rho_z$ is defined as $\rho_z \equiv \int {v_i^2|\psi_{\nu\pi m}|^2 dxdy}$ with $v_i^2 (\in[0,2])$ the occupation number of the orbit. Density distributions for PKO1 and PKDD exhibit significant accumulations beyond $|z| > 3$~fm, with integrated densities exceeding the critical 4 fm$^{-3}$ threshold, suggesting $\alpha$-clustering formation. The density of PKO1 peaks at $z \approx 4.0$ fm, while the density of PKDD peaks at $z \approx 1.5$ fm. The cluster-forming regions originate from specific orbitals. In the excited state, cluster-forming regions involve $1/2_1^-$, $1/2_2^-$, and $1/2_2^+$ orbitals, contrasting the ground-state $^{20}$Ne configuration ($1/2_1^-$ + $1/2_2^+$). In particular, the $1/2_2^-$ orbital (solid red curve) transitions from oblate to prolate geometry during the evolution of quadrupole deformation. These single-particle levels have the characteristic of $m = 1/2$ and are accompanied by parity mixing. This has also been confirmed in the 4-$\alpha$ chain states of $^{16}$O, and the reason may be that the $\alpha$ particles are $0^+$ bosons. 
	
	The third column of Fig.\ref{fig:3} illustrates the situation of PKO1 without $V_{\pi-PV}^T$. The contour plots of density distribution are similar to Lagrangians PKO1 and PKDD in the ground state. In particular, under large prolate deformation, the absence of additional attraction from the tensor force component of $\pi$-PV causes the minimum of $E_B$ to converge at a quadrupole deformation smaller than that of the full PKO1 Lagrangian. Furthermore, unlike PKO1 which shows a density peak at $z \approx 4$~fm, PKO1 without $V_{\pi-PV}^T$ shows a density peak at $z \approx 1.5$~fm. The results obtained with the effective Lagrangian PKO1 show a maximum density in the clustering region, indicating that the $\pi$-PV coupling has an effect on the compactness of the $\alpha$ cluster.
	
	\begin{table}[h!]
		\caption{The characteristic quantities to describe the nucleon localization and clustering in $^{20}$Ne, obtained by the D-RHFB calculation with Lagrangians PKO1, PKO2, PKO3 and PKDD. For comparison, the results with PKO1 but excluding tensor force component of $\pi$-PV contributions from the nucleon self-energy (PKO1 w.o. $V_{\pi-PV}^T$) is given as well. The criterion values are taken from Ref. \cite{PhysRevC.106.064330}.}
		\label{tab:table2}
		\begin{tabular}{lcrrrr}
			\toprule
			&  &${C_{q\sigma }}$                  &          $\rho /{\rho _{\max }}$      &                    $F{({R_\alpha })_ \parallel }$       &           $F{({R_\alpha })_ \bot }$ \\
			\hline

			\multirow{2}{*}{PKO1} & E.S. & 0.99 & 1.00 & 0.78 &0.85\\
			& G.S. & 0.99 & 1.00 & 0.79 &0.74\\
			\hline
			PKO1 & E.S. & 0.99 & 0.90 & 0.76 &0.84\\
			w.o. $V_{\pi-PV}^T$ & G.S. & 0.99 & 1.00 & 0.73 &0.70\\
			\hline
			\multirow{2}{*}{PKO2} &E.S.  & 0.99               & 0.93             &       0.75         &       0.84\\
			& G.S. &0.99                 &   1.00              &         0.71        &        0.65\\
			\hline
			\multirow{2}{*}{PKO3} & E.S.  &  0.99               &     1.00              &         0.77         &       0.84\\
			& G.S. &0.99                  &  1.00                &       0.78          &     0.73\\
			\hline
			\multirow{2}{*}{PKDD} & E.S.  &  0.99               & 0.94               &     0.76             &   0.83\\
			& G.S. &0.99                 &   1.00                 &      0.75            &    0.69\\
			\hline
			Criteria \cite{PhysRevC.106.064330}     &     &   $ >0.90  $      &          $ >0.80  $        &         $  >0.60 $      &       $ >0.60$\\
			\bottomrule
		\end{tabular}
	\end{table}
		
	In order to extract the clustering signatures of $^{20}$Ne with a superdeformed prolate, we analyze the wave function of $^{20}$Ne using the nucleonic localization function and $F$ factor. Table.\ref{tab:table2} presents the calculated results of the NLF and $F$ factor for Lagrangians PKO$i$ and PKDD. $C_{q\sigma }$ represents the result of NLF, with ${C_{q\sigma }}$ close to 1 being a necessary condition for the existence of an $\alpha$ cluster. For $F$ factor quantification requiring extraction of density maxima, the clustering center positions are determined at $z \approx 3$ fm (ground state) and $z \approx 4$ fm (excited state), corresponding to the respective density peaks, where the radius of $\alpha$ cluster ($R_\alpha$) is selected as 1.9 fm. The $\rho / \rho_{\max}$, $F{({R_\alpha })_ \parallel }$, and $F{({R_\alpha })_ \bot }$ are indicators of the compactness of the cluster. The criterion for determining the existence of $\alpha$ cluster is taken from Ref. \cite{PhysRevC.106.064330}. Table.\ref{tab:table2} shows that both PKO$i$ and PKDD Lagrangians indicate the presence of $\alpha$ clusters in both ground and excited states. In the excited state, the calculated results of $F{({R_\alpha })_ \parallel }$, and $F{({R_\alpha })_ \bot }$ are larger than those in the ground state, implying a faster density drops and enhanced compactness of $\alpha$ Cluster. The results for PKO1 are stronger than those for PKO2 and PKDD, suggesting that the $\pi$ -PV coupling enhances the compactness of the clustering. For comparison, Table.\ref{tab:table2} also includes results for PKO1 without $V_{\pi-PV}^T$. Calculations  for PKO1 without $V_{\pi-PV}^T$ still show the existence of $\alpha$ clustering, but its $F$ factor is significantly reduced. This demonstrates that the tensor force component of $\pi$-PV coupling enhances the density gradient at $\alpha$ cluster boundaries, thereby enhancing the compactness of $\alpha$ clusters.

	\begin{figure}[t]
		\centering
		\includegraphics[width=\columnwidth]{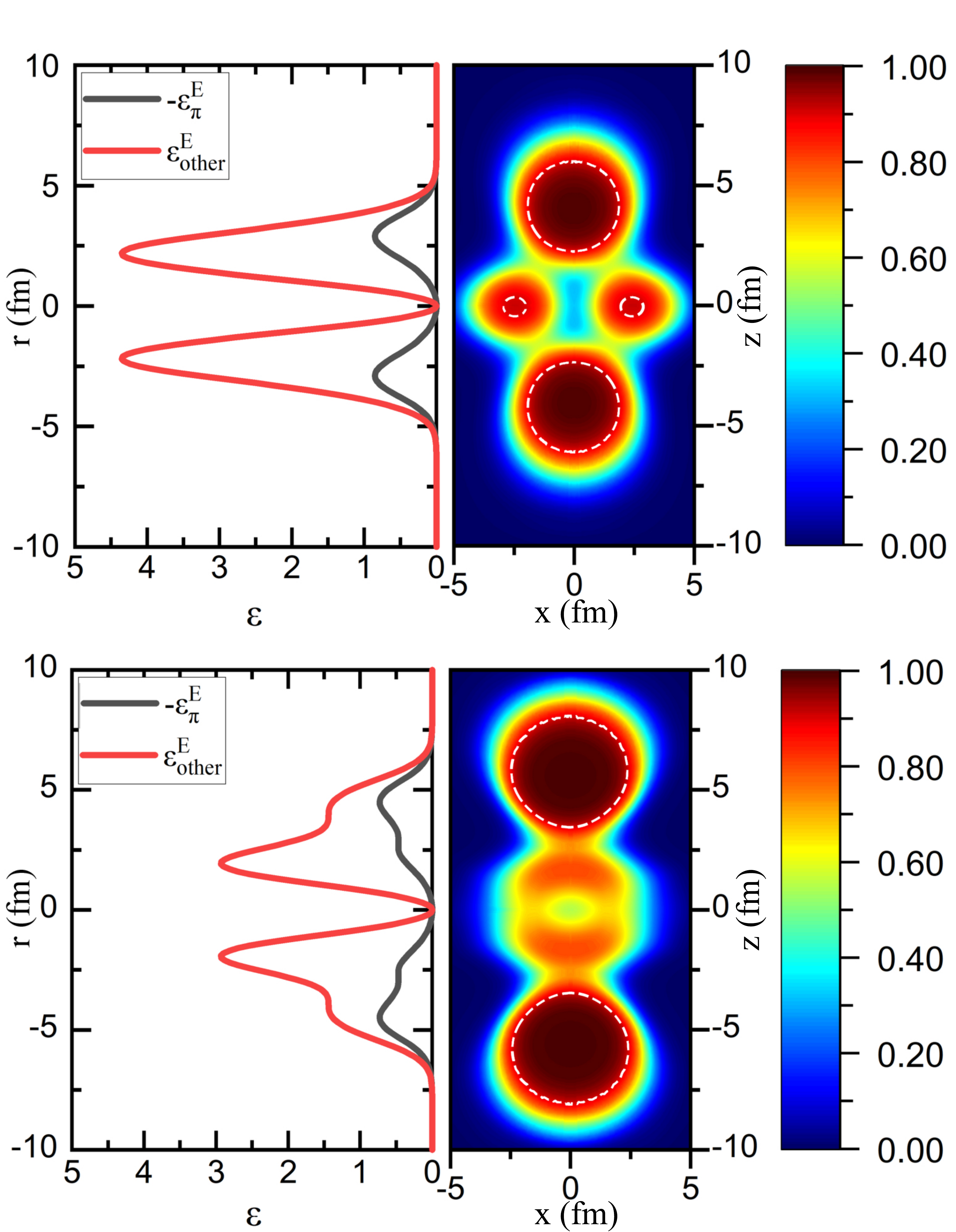}
		\caption{The energy functional integral element $\varepsilon$ of Fock terms as a function of $r$ distribution (left figure) and the contour plot of the nucleonic localization function (right figure), both are calculated by the D-RHFB model with Lagrangians PKO1. The upper figures represent the ground state of $^{20}$Ne, while the lower figures represent the excited state. The gray line represents the distribution of the energy functional integral element for $\pi$-PV coupling with respect to $r$, while the red line represents the distribution for the sum of the coupling channels $\sigma$, $\omega$, and $\rho$. The distribution of $\varepsilon$ reveals the strong attractive effect of the $\pi$-PV in the $\alpha$ cluster region ($z\approx$4~fm) and the mechanism of its long-range cancellation of Coulomb repulsion. The while dash lines indicate the region of ${C_{q\sigma }\geq 0.9}$.} 
		\label{fig:4}
	\end{figure}
	
	To further extract the impact of pion meson coupling channel on the $\alpha$ cluster, we define the energy functional integral element $\varepsilon$ for each coupling channel, satisfying
	\begin{align}
		E_\phi ^E = \int {{\varepsilon _\phi^E } dr} 
	\end{align}
	where $\phi$ represents $\sigma,\omega,\pi $ and $\rho$. Fig.\ref{fig:4} shows the relationship between $\varepsilon _\phi^E$ and $r$ for different coupling channels. Note that the value of $\varepsilon _\pi^E$ is actually negative, which means that $\pi$-PV provides an attractive effect. Moreover, since the variation trends of the coupling channels other than $\pi$ are the same, we combined them with the subscript 'other' including $\sigma$, $\omega$, and $\rho$. 
	
	In Fig.\ref{fig:4} the contour plot of the NLF suggests the existence of $\alpha$ cluster in $^{20}$Ne in both the ground state and the excited state. Compared with density profiles, the localization function generally exhibits greater spatial extent due to the inclusion of the kinetic energy term in its formulation. The nucleonic localization function reveals pronounced localization maxima at the outer extremities and over the alpha clusters in $^{20}$Ne. This spatial arrangement strongly supports an $\alpha$-$^{12}$C-$\alpha$ quasimolecular configuration~\cite{ZHOU2016227}. In the ground state, the distributions of $\varepsilon_\pi^E$ and $\varepsilon_{\mathrm{other}}^E$ as a function of $r$ are similar, whereas in the excited state, they differ significantly. As demonstrated in Fig. \ref{fig:4}, the $\pi$-PV coupling manifests significantly stronger attractive interactions near $\alpha$-cluster regions than in the vicinity of the $z=0$ plane. When inter-cluster distances exceed 7~fm, Coulomb repulsion becomes the predominant interaction mechanism~\cite{PhysRevC.69.024309}, while the long-range $\pi$-PV attraction effectively counteracts this repulsive force. This phenomenon elucidates the microscopic mechanism through which $\pi$-PV increases the binding energy in systems containing $\alpha$ clusters. Comparison of Fig. \ref{fig:3} (d,f) reveals enhanced cluster formation under the PKO1 functional, influenced by the $\pi$-PV coupling. This interaction critically balances Coulomb repulsion to maintain a spatial separation between clusters and the nuclear core, thereby effectively preventing excessive dissolution of clusters within the nuclear medium.
	
	The role of pion tensor force is further elucidated via DWS basis analysis. Table.\ref{tab:table3} shows the squared amplitudes $|C_{\nu \pi m}|^2$ of spherical components $l_j$ in the deformed single-particle levels $m_{\nu}^\pi$ for both ground state and excited state. In the ground state, the negative parity state that provides the clustering density is ${1/2}_1^-$, which is almost entirely composed of $p$ spin-partner states. When it comes to excited states, the negative parity states that contribute to the clustering density are ${1/2}_1^-$ and ${1/2}_2^-$. For Lagrangians PKO$i$ and PKDD, ${1/2}_1^-$ both have approximately 80\% $p$ state components and 15\% $f$ state components, while ${1/2}_2^-$ has about 60\% $p$ components and 30\% $f$ components. An angular momentum difference $\Delta L = 2$ between the DWS basis in the deformed single-particle energy levels leads to an enhancement of the $\pi$-PV coupling~\cite{PhysRevC.101.064302,PhysRevC.105.034329}. The tensor force-induced coupling between $pf$ orbitals of the spherical DWS basis in the single-particle energy levels ${1/2}_1^-$ and ${1/2}_2^-$ reduces their single-particle energies and enhances deformation stability. A similar mechanism applies to the ${1/2}_2^+$ level, but with orbital coupling shifting to $sd$ orbitals of the spherical DWS basis. At the same time, due to the low population of $f$ components in the ground state ${1/2}_1^-$ orbit, there is no enhancement of the $\pi$-PV coupling observed. This explains how $\pi$-PV coupling stabilizes nuclei under large prolate deformations.
	
	\begin{table}[h!]
		\caption{The percentage of spherical components $l_j$ in the deformed single-particle energy levels $m_{\nu}^\pi$ for both ground state (G.S.) and excited state (E.S.) of $^{20}$Ne, calculated by the D-RHFB model with Lagrangians PKO1. For comparison, the results with tensor force component of $\pi$-PV contributions removed from the nucleon self-energy are given as well.}
		\label{tab:table3}%
		\begin{tabular}{cccccccccccc}
			\toprule
			\multirow{2}{*}{$m_{\nu}^\pi$} & \multirow{2}{*}{$l_j$} &\multicolumn{2}{c}{PKO1}  &\multicolumn{2}{c}{PKO1 w.o. $V_{\pi-PV}^T$} \\
			\multicolumn{2}{c}{ }&E.S.& G.S. &E.S.& G.S. \\
			
			\midrule
			
			\multirow{2}{*}{${1}/{2}_1^ + $}& $s_{{1}/{2}}$ &0.934	&0.984	&0.963&0.985\\
			
			& $d_{{3}/{2}}$  +$d_{{5}/{2}}$  &0.058	&0.014	&0.032&0.014\\
			
			\midrule
			
			\multirow{2}{*}{${1}/{2}_2^ + $}& $s_{{1}/{2}}$  &0.273	&0.188		&0.274&0.192\\
			
			& $d_{{3}/{2}}$  +$d_{{5}/{2}}$   &0.549 & 0.778 &0.574&0.779  \\
			
			\midrule
			
			\multirow{2}{*}{${1}/{2}_1^ - $}& $p_{{1}/{2}}$+$p_{{3}/{2}}$ & 0.815&0.961&0.830&0.971\\
			
			& $f_{{5}/{2}}$  +$f_{{7}/{2}}$      & 0.156 & 0.037 &0.146& 0.028 \\	
			
			\midrule
			
			\multirow{2}{*}{${1}/{2}_2^- $}& $p_{{1}/{2}}$+$p_{{3}/{2}}$ &0.592 &0.993&0.663&0.993\\
			
			&$f_{{5}/{2}}$  +$f_{{7}/{2}}$     & 0.304   & 0.002  & 0.263&0.006\\
			
			\midrule
			
			\multirow{2}{*}{${3}/{2}_1^ - $}& $p_{{3}/{2}}$ & 0.950 & 0.994  &0.965&0.992\\
			
			& $f_{{5}/{2}}$  +$f_{{7}/{2}}$      & 0.045  & 0.005  &0.031&0.007  \\
			\bottomrule  	    
		\end{tabular}
	\end{table}
	
	In summary, clustering dynamics in prolate-deformed $^{20}$Ne are investigated using the D-RHFB framework with PKO$i$ and PKDD Lagrangians. The D-RHFB model can self-consistently handle tensor forces, deformation, pairing correlations and continuum effects, offering a useful tool for analyzing clustering features likely to emerge under extreme large deformation. Combining NLF and $F$-factor analyses, we systematically unravel critical role of $\pi$-PV coupling in $\alpha$-cluster formation and stabilization of prolate-deformed $^{20}$Ne. The tensor force of $\pi$-PV coupling exhibits three pivotal mechanisms: (1) It induces $sd$ ($pf$) orbital mixing in the canonical single particle levels with positive (negative) parity, enlarging splittings from degenerate spherical orbits; (2) It alters the shell closure structure and increases the stability of the $^{20}$Ne in the superdeformed prolate state. (3) It sharpens $\alpha$ cluster localization and suppresses $\alpha$ cluster dissolution through a delicate balance between long-range $\pi$-PV attraction and Coulomb repulsion. This study reveals the effects of the tensor force on nuclear structure and clustering in light nuclei, offering insights into the understanding of quantum phase transition in nuclear many-body systems.
	
	Notably, while the current study provides critical insight into the effects of tensor force on clustering in the $\alpha+{^{12}}{\rm C}+\alpha$ configuration, our D-RHFB framework remains limited by the omission of beyond-mean-field corrections~\cite{PhysRevLett115-022501, zhou_anatomy_2016}. Considering correlations beyond the mean field may further reduce the energy of the $^{20}$Ne nucleus in the superdeformed prolate state and narrow the gap between its energy and the 2-$\alpha$ threshold. Future efforts could prioritize integrating collective correlation mechanisms to refine excitation energy calculations and enhance agreement with experimental data. Furthermore, the present RHF framework is restricted to axially symmetric shapes, which precludes investigations of octupole-deformed systems~\cite{CPC:10.1088/1674-1137/adc11e}. In octupole-deformed systems, the tensor force may investigated for its role in clustering phenomena, such as in the $\alpha+{^{16}}{\rm O}$ configuration. The expansion of research scope will deepen our understanding of clustering mechanisms driven by nuclear in-medium force and their interplay with nuclear deformation.
	
	\section*{Acknowledgement}
	The authors are grateful to Dr. Jing Geng for his helpful discussion. This work was partly supported by the National Natural Science Foundation of China (11875152), the Strategic Priority Research Program of Chinese Academy of Sciences (XDB34000000), and the Fundamental Research Funds for the Central Universities, Lanzhou University (lzujbky-2023-stlt01). The authors also want to thank the computation resources provided by the Supercomputing Center of Lanzhou University.
	
	\bibliographystyle{elsarticle-num}

\end{document}